\documentstyle[aps,graphicx]{revtex}
 \def\ket{\rangle}
\draft 
\def\<{\langle}
\def\>{\rangle}
\begin{document}

\title{An Electron-Nucleon  Double Spin Solid-State Quantum Computer }
\author{Gui Lu Long$^{1,2,3,4}$, Ying Jun Ma$^{1,2}$, and Hao-Ming
Chen$^1$ }

\address{
$^1$Department of Physics, Tsinghua University, Beijing 100084, China\\
$^2$ Key Laboratory For Quantum Information and Measurements,
Beijing 100084, China\\
$^{3}$ Center for Atomic and Molecular Nanosciences, Tsinghua
University, Beijing 100084, China\\
$^4$ Institute of Theoretical Physics, Chinese Academy of
Sciences, Beijing 100080, China}
\date{\today}
\maketitle
\begin{abstract}
An electron-nucleon double spin(ENDOS) solid-state quantum
computer scheme is proposed. In this scheme, the qubits are the
nuclear spins of phosphorus ion implanted on the (111) surface of
$^{28}$Si substrate. An $^{13}$C atom
 on a scanning tunnelling probe tip is used both to
complete single qubit and two-qubit control-not operation, and single qubit measurement.
The
scheme does not require interactions between qubits, and can
accomplish two qubits without the use of SWAP gate.
 This scheme is scalable, and
can be implemented with present-day or near-future technologies.
\end{abstract}


Quantum computer(QC) is a worldwide scientific
pursuit.  To design a QC, one  needs  a two level
quantum system to act as qubit and able perform single bit
rotation, two qubit control-not operation and single qubit measurement.
The system should
also be robust against errors and decoherence. Many realization
schemes have been put forward, for instance, liquid nuclear
magnetic resonance\cite{r1,r2}, ion trap\cite{r3},
cavity-QED\cite{r4}, Josephson-junction qubit\cite{r5,r5p},
quantum dot\cite{r6}, nuclear spin qubit in solid state and liquid crystal\cite{r7},
nuclear spin qubit in liquid crystal lattice\cite{r8}, photon
qubit\cite{r9}, electron qubit\cite{r10}, magnetic force scanning
microscopy\cite{r12} are just a few names in this long list.

Nuclear spins are well isolated from environment, their coherence
time is usually very long at low temperature\cite{r7}. Meanwhile,
modern micro-fabrication technology is well developed,
making solid state device a favorite choice. In Kane's proposal\cite{r7},
these two advantages are combined.
Logical operations are completed by using externally applied
electric field and radio frequency pulses.
One important feature of Kane's
scheme is that it requires atomic size precision of the donor
phosphorus ion, and delicate design of the spin measurement which
is a feast for the present-day technology.
Several alternative proposals have
been put forward that do not need such high precision
requirement,  using scanning tunneling microscope(STM) or magnetic
resonance force microscopy(MRFM)\cite{r12,r17},
nuclear spins at lattice structure\cite{r8,r18}. Common in these
schemes, are the use of gradient magnetic fields and the
dipole-dipole interaction between neighboring qubits. In a
gradient magnetic field, different qubits have different
frequency so that it can be individually addressed. The
(dipole-dipole) interaction between neighboring qubits is
essential to construct the CNOT operation in these schemes.  However,
additional effort has to be made to cancel these dipole interaction and the next
neighbor interaction when they are not used in the computation.

STM has been broadly used in various fields of research since its
invention\cite{r18}. It can perform single atom engineering
\cite{r20,r22,r23,r24,r25}. STM can manipulate a single atom or a
molecule\cite{r20,r26,r27}, and even doing so at room
temperature. The spatial resolution of STM has
reached the sub-{\AA}ngstr\"{o}m range\cite{r28}. In particular,
recent  studies have demonstrated the
possibility of detecting single spin state using the modulation
effect of STM tunneling current\cite{r29,r29p,r30,r32p}. This
makes the STM a very good candidate for quantum computer design.

\begin{figure}
\begin{center}
\caption{An illustration of the ENDOS scheme. A $^{13}$C is
placed at the STM tip. The qubits are the nuclear spins of
$^{31}$P ion deposited on the surface of $^{28}$Si substrate.
When the STM tip approaches an ion, it  changes the contact
interaction and resonance frequency.}
\includegraphics[width=5cm,angle=0]{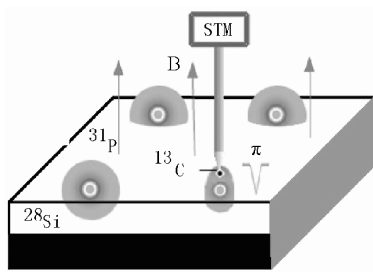}
\label{f1}
\end{center}
\end{figure}

Here we propose an electron-nucleon double spin(ENDOS) solid state
quantum computer that uses STM for single qubit addressing, operation,
two-qubit operation and final state measurement. The ENDOS
scheme is shown in Fig.{\ref{f1}}. Phosphorus ion arrays are
deposited at the (111) surface of $^{28}$Si substrate. Over the
sample there is an STM tip which can move vertically and
horizontally. The whole set-up is placed in an electron/nucleon
magnetic resonance chamber where a strong static magnetic field
$B$ along $z$-axis is provided, and the radio frequency(r.f.)
waves for electron/nucleon resonance manipulation is applied. The
STM tip is made of carbon nano-tube with a $^{13}$C at the tip.
This can be accomplished using present-day technology. Carbon
nanotube can be conductive, it is very good for STM purpose. The
donor {\it\bf $^{31}$P has nuclear spin} $I=1/2$, and it is used
as {\it\bf qubit}. The electrons of the phosphorus ions are weakly
bound at low temperature, and they are used in manipulating the
qubit. The distance of $^{31}$P ions in the substrate can be very
large, say $a\ge 30nm$, since we do not require the interaction
between neighboring qubits. This brings two benefits in quantum
computation. First, there is not atomic-size precision
requirement in fabrication. Second, ENDOS scheme is free from the
trouble to decouple the dipole-dipole interaction between
neighboring qubits when they are not in operation. Further the
interaction of qubit between the next neighbor qubits cause
considerable problems and this is also avoided in the ENDOS
scheme.

The setup is placed at low temperature. In the presence of the
magnetic filed $B$, both nucleon and electron spin levels undergo
Zeeman splitting. Because the gyromagnetic ratio of electron is
negative, $g_e=-2$ for electron, spin down is its ground state.
The ground state of phosphorus ion and $^{13}$C are spin up,
because $g_n^p=2.26$ and $g_n^a=1.4048$ respectively. We label
their ground states as $|0\ket$. Before a quantum computation
starts, one must first {\it\bf initialize all the qubits into the
ground state}. This is done in two steps. First at a  temperature
of 1 $K$,  all the electrons of the $^{31}$P ions are in their
ground state and half of the $^{31}$P nuclear spins are in their ground state.
Then we use the STM tip to measure the state of the
$^{31}$P nuclear spin, if it is in the excited state, a  nuclear
$\pi$ pulse is applied to that qubit to flip its spin. In this
way, the quantum computer can be initialized. Details will be
given shortly for single qubit  measurement.

 {\it \bf Single qubit
addressing} is completed by moving the STM tip toward the qubit.
The voltage of the STM tip changes the hyperfine interaction of
the qubit, and hence the NMR frequency. When the STM tip is
lifted up, the Hamiltonian of the nuclear spin -electron system
of an ion can be written as\cite{r31}
\begin{equation}
H_{e-p}=g_{e}\mu
_{B}BS_{z}-g_{n}^{p}\mu_{n}BI_{z}^{p}+hA_{z}S_{z}I_{z}^{p}.
\label{e4}
\end{equation}
where $hA_{z}$ is the hyperfine coupling energy between $^{31}$P
ion and the electron and its value can be measured by NMR
experiment($A_z=120$MHz was given in Ref.\cite{r32}), $g_{n}^{p}$
is the Land\.{e} $g$-factor of $^{31}$P ($g_n^p\simeq 2.26$) and
$g_e=2$, $\mu_B$ and $\mu_n$ are Bohr and nuclear magneton
respectively. When we move the STM tip close to this ion, say at
a distance of 10 {\AA}ngstr\"{o}m,  the tip-electron-ion system
Hamiltonian becomes
\begin{eqnarray}
H_{a-e-p}&=&H_{e-p}'-g_{n}^a\mu
_{n}BI_{z}^{a}+hA^{^{\prime}}S_{z}I_{z}^{a}\nonumber\\
&=& g_{e}\mu
_{B}BS_{z}-g_{n}^{p}\mu_{n}BI_{z}^{p}+hA_{z}'S_{z}I_{z}^{p}-g_{n}^a\mu
_{n}BI_{z}^{a}+hA^{^{\prime}}S_{z}I_{z}^{a} \label{e5}
\end{eqnarray}%
where $I_{z}^{a}$ is the nuclear spin of the tip $^{13}$C, its
gyromagnetic ratio is $g_{n}^{a}\simeq 1.4048$. The $hA'_Z$ is
the hyperfine interaction between $^{31}$P ion and the electron
spin. A prime indicates that its value has been  modified by the
presence of the STM tip. $hA'$ is the hyperfine interaction
between the electron and the $^{13}$C ion where a prime also
indicates a modification due to the presence of the phosphorus
ion. Their exact values should be determined through experiment.
The value of $A'$ is of the order of a few GHz\cite{r12}. The
value of $A'_z$ is in the order of MHz. Without the STM tip
modification, it was reported that $A_z=120$MHz\cite{r32}.

 To perform a {\bf single qubit rotation}, we move the STP tip
close to the qubit, say, 10 {\AA}ngstr\"{o}ms. The $^{13}$C
nuclear spin and the electron spin are in ground state.
 The $^{31}$P nuclear spin resonance
frequency is now
\begin{eqnarray}
\hbar\omega_{i}'=g_n^p\mu_n B-\hbar A'_z/2,
\label{1qubit}
\end{eqnarray}
which is different from that in the absence of the STM tip
$\hbar\omega_{i}=g_n^i\mu_n B-\hbar A_z/2$. Here the STM tip acts
just like the $A$-gate in Kane's proposed quantum
computer\cite{r7}. When  a r.f. pulse with frequency at (\ref{1qubit})
is applied to the sample,  the qubit under the STM tip is in resonance and
undergo appropriate rotation just like in a NMR microscopy.

The biggest appeal of STM is the possibility of performing single qubit detection.
{\it\bf Single nuclear spin detection} can be done by the
modulation of nuclear spin on the STM tunneling current. When the STM tip is close
to a magnetic center, such as an electron spin,
the STM tunneling current is modulated by the magnetic center's state. By
 analysing the spectrum of the  tunneling
current, one can determine the state of the qubit. The modulation
effect of a localized electron spin to the tunneling current has
already been observed in Si/Si$_2$\cite{r32p}, Fe \cite{r29}, and
very recently in BDPA molecule\cite{r29p}. Though there is no
direct experiment of the single spin detection of $^{31}$P, we
believe that present technology is close to this purpose. As in
Ref.\cite{r12}, we assume that single qubit detection can be
realized in this way.  Here we assume the modulation of tunneling
current is independent of the conducting electron spin
polarization as suggested in Ref.\cite{r28}, though
spin-dependence mechanism is also suggested\cite{r32ppp}. Nuclear
spins can influence the electron through dipole interaction and
thus causes a change in the modulation frequency.
 Both the STM $^{13}$C
nuclear spin and the phosphorus nuclear spin have modulation
effects on the STM electron tunneling current.
The
electron tunneling current frequency is modulated by the $^{31}$P
nuclear spin and the $^{13}$C nuclear spin in the following way:
\begin{eqnarray}
f_{e,p=0}&=g_e\mu_B B\mp{1\over 2} h A'+ h A'_Z/2,\;\;\;\;\;
&^{31}{\rm P
\;\; nuclear\;\; up}\label{modul0}\\
f_{e,p=1}&=g_e\mu_B B\mp{1\over 2} h A'-h
A'_Z/2,\;\;\;\;\;\;\;&^{31}{\rm P \;\;nuclear\;\; down},\label{modul1}
\end{eqnarray}
where the plus(minus) sign in front of $hA'/2$ corresponds to $^{13}$C
in $|0\ket$ state($|1\ket$ state).

{\bf Two-qubit control-not operation} construction is vital in a quantum computer scheme.
Though there are many possible unitary operations in a quantum computation. It has been
shown that all the unitary operations can be constructed by combining single qubit rotation and
two-qubit CNOT operation. If these gates are constructed, then a quantum computer structure is
constructed. We require the STM tip and all the electrons remain in ground states before and
after the gate operation so that follow up gate operations using the same tip and
electron condition.  A CNOT gate is a conditional NOT gate.
 Suppose at the start, the control qubit state is
 $\alpha'|0\ket_c +\beta' |1\ket_c$ and the target qubit  state
is $\alpha|0\ket_t +\beta|1\ket_t$. A CNOT
operation  yields the following state: $\alpha'|0\ket_c
(\alpha|0\ket_t+\beta|1\ket_t)$
$+\beta'|1\ket_c(\alpha|1\ket_t+\beta|0\ket_t$ where the
subscripts $c$ and $t$ refer to control- and target- qubits
respectively. We use $ec$ and $et$ subscripts to indicate the state of the electrons of
the control and target ions. Unlike in schemes that use neighboring
dipole-dipole interaction where direct CNOT gate can only be
constructed for neighbors, the control-qubit and the target-qubit
in ENDOS scheme need not be neighbors. They can be any distance
away. The CNOT gate implementation is implemented through  7
steps. The procedures involve first combined electron
 and nuclear resonant pulses to entangle the control qubit
with its electron, then with the tip nuclear spin whose state is labeled by subscript $a$,
and then with the target qubit electron, and then the target qubit.
 Then another
combined sequence of pulses disentangle the electrons and the
$^{13}$C nuclear spin from the two qubits involved. As can be seen from
equation (\ref{e5}), the nuclear spin resonance is only modulated
by the electron spin state, whereas the electron resonance
frequency is modulated by both the $^{13}$C nuclear spin and the
$^{31}$P nuclear spin. The 7 steps for a CNOT gate operation are

1) Move the STM tip to the control qubit so that its resonance frequency is changed.
Apply an electron
$\pi$-pulse with frequency
\begin{eqnarray}
f_{ec}=g_e\mu_eB+{hA'\over 2}-{hA'_Z\over 2}. \label{fec}
\end{eqnarray}
This pulse sequence changes the state of control-qubit electron if the control qubit
is in state $|1\ket$.  The effect is
\begin{eqnarray}
(\alpha'|0\ket_c+\beta'|1\ket_c)|0\ket_{ec}\rightarrow (\alpha'|0\ket_c|0\ket_{ce}
+\beta'|1\ket_c|1\ket_{ec}).
\end{eqnarray}
Thus it entangles the control qubit with its electron spin. This
is already within the current technology. Entanglement of nuclear
spin and electron spin  at $T=40$K for radical CH has just been
realized experimental by Mehring et al\cite{r32pp}.

2) Apply a nuclear $\pi$-pulse with frequency
\begin{eqnarray}
f_a=|g^a_n\mu_N B-{h A'\over 2}|. \label{fa}
\end{eqnarray}
This frequency changes the state of STM tip $^{13}$C nuclear spin on condition
that the control electron is
in state $|1\ket$. After this pulse, the STM tip also joins the entanglement. The wave
function goes through the following change
\begin{eqnarray}
(\alpha'|0\ket_c|0\ket_{ce}+\beta'|1\ket_c|1\ket_{ec})|0\ket_a)
\rightarrow \alpha'|0\ket_c|0\ket_{ce}|0\ket_a+\beta'|1\ket_c|1\ket_{ec}|1\ket_a).
\end{eqnarray}

3) Move the STM tip to the target ion and  apply two electron
$\pi$-pulses so that the electron spin of the target ion is
flipped on condition that the STM tip is in state $1$.  Quantum mechanics ensures that
even though the STM tip has been shifted elsewhere, the entanglement of the control
qubit and its electron and the STM tip still holds.
The two pulse frequencies are
\begin{eqnarray}
f_{et1}= g_e\mu_B B+{hA'_Z\over 2}+{hA'\over 2},
f_{et0}= g_e\mu_B B-{hA'_Z\over 2}+{hA'\over 2}\label{fet0}
\end{eqnarray}
respectively. They  change the system into state
\begin{eqnarray}
\alpha(\alpha'|0\ket_t+\beta'|1\ket_t)|0\ket_{et}|0\ket_c|0\ket_{ec}|0\ket_a
+\beta
(\alpha'|0\ket_t+\beta'|1\ket_t)|1\ket_{et}|1\ket_c|1\ket_{ec}|1\ket_a.
\end{eqnarray}

4) Apply a nuclear $\pi$-pulse on the target qubit with frequency
\begin{eqnarray}
f_{p}=g^p_n\mu_N B-{h A'_Z\over 2}.
\end{eqnarray}
This pulse change the system into state
\begin{eqnarray}
\alpha(\alpha'|0\ket_t+\beta'|1\ket_t)|0\ket_{et}|0\ket_c|0\ket_{ec}|0\ket_a
+\beta
(\alpha'|1\ket_t+\beta'|0\ket_t)|1\ket_{et}|1\ket_c|1\ket_{ec}|1\ket_a.
\end{eqnarray}
Thus up to now, we have implemented the CNOT operation on the two
qubits, but entangled with the two electrons and the STM tip.
Next we disentangle the electrons and the STM tip from the
control- and target- qubits.

5) Apply two electron $\pi$-pulses on the target ion electron with
frequencies $f_{et1}$ and $f_{et0}$ respectively as given in
Eqns. (\ref{fet0}). These two electron pulses
disentangle the target ion electron. The system state becomes
\begin{eqnarray}
\alpha(\alpha'|0\ket_t+\beta'|1\ket_t)|0\ket_{et}|0\ket_c|0\ket_{ec}|0\ket_a
+\beta
(\alpha'|1\ket_t+\beta'|0\ket_t)|0\ket_{et}|1\ket_c|1\ket_{ec}|1\ket_a.
\end{eqnarray}

6) Move the STM tip back to the control qubit. Apply a nuclear
$\pi$-pulse on the STM tip by using frequency $f_a$ as given in
(\ref{fa}). This disentangles the STM tip $^{13}$C nuclear spin.
The system state becomes
\begin{eqnarray}
\alpha(\alpha'|0\ket_t+\beta'|1\ket_t)|0\ket_{et}|0\ket_c|0\ket_{ec}|0\ket_a
+\beta
(\alpha'|1\ket_t+\beta'|0\ket_t)|0\ket_{et}|1\ket_c|1\ket_{ec}|0\ket_a.
\end{eqnarray}

7) Apply an electron $\pi$ pulse on the control ion electron by
using frequency $f_{ec}$ as given in (\ref{fec}). This
disentangles the control ion electron. Then finally the state of
the system is
\begin{eqnarray}
\left\{\alpha(\alpha'|0\ket_t+\beta'|1\ket_t)|0\ket_c +\beta
(\alpha'|1\ket_t+\beta'|0\ket_t)|1\ket_c\right\}|0\ket_{et}|0\ket_{ec}|0\ket_a.
\end{eqnarray}

The speed of quantum computation in ENDOS depends on the duration of eletron/nucleon pulse
length and the time the STM tip takes to move from one position to another position.
Typical nuclear pulse is of the order of micro-second and electron pulse is much shorter.
The main factor is the STM tip speed. It is estimated that current commercial STM machine
can make 8 scans per second on an area of 10nm$\times$10nm with 258$\times$258 points in each
scan, which is equivalent to scan a point in 15$\mu$s. Thus the estimated time for a CNOT gate
is the order of $100\mu$s. If the decoherence time is 10 s, then it is sufficient to finish
$10^5$ gate operations. The speed can be accelerated if we make several STM tips
working in parallel.

To summarize, an electron-nucleon double spin quantum computer
scheme is proposed. Basic gate operations and measurement
 can be implemented by a STM tip with a $^{13}$C.
The
scheme is within or close to the reach of present-day technology.

Helpful discussions with Prof. Kuzmany and Dr. Markus Arndt of University of Vienna,
Dr Ille Gebeshuber of TU Vienna are gratefully acknowledged.
We also thank Ms Li Xiao and Ms Fang Liu for help.
This work is supported in part by China National Science
Foundation, the National
Fundamental Research Program, Contract No. 001CB309308 and the
Hang-Tian Science foundation.

\end{document}